\documentclass[12pt]{article}

\usepackage[dvips]{graphicx}
\usepackage{epsfig}
\usepackage{amsmath,amsfonts,amssymb,amsthm}
\usepackage{mathrsfs,mathtools}
\usepackage{verbatim}
\usepackage{psfrag}
\usepackage{bm}
\usepackage[english]{babel}
\usepackage{bbm}
\usepackage[utf8]{inputenc}
\usepackage[square,comma,sort&compress,numbers]{natbib}
\usepackage[dvipsnames]{xcolor}
\usepackage{slashed}
\usepackage{upgreek}
\usepackage{enumitem}
\usepackage{float}
\usepackage[T1]{fontenc}
\usepackage{soul}
\usepackage{pgfplots}
\usepackage{extarrows}
\usepackage{appendix}
\pgfplotsset{compat=1.18}
\usetikzlibrary{external}
\usepackage{array}

\usepackage[T1]{fontenc}

\usepackage[colorlinks=true,urlcolor=blue,anchorcolor=blue,citecolor=blue,filecolor=blue
,linkcolor=blue,menucolor=blue,linktocpage=true,pdfproducer=medialab,pdfa=true
]{hyperref}
\usepackage{epsf,epsfig}
\usepackage{graphics}
\usepackage{subcaption}
\usepackage{moresize}
\usepackage{cancel}

\newcommand{\cb}{\color{black}}
\newcommand{\cbl}{\color{black}}
\newcommand{\bL}{\begin{Large}}
\newcommand{\eL}{\end{Large}}
\newcommand\blfootnote[1]{%
  \begingroup
  \renewcommand\thefootnote{}\footnote{#1}%
  \addtocounter{footnote}{-1}%
  \endgroup
}

\def\vec{\mathbf}

\def\P{\mathcal{P}}

\def\sech{\rm Sech}

\def\half{\textstyle{\frac{1}{2}}}
\def\fourth{\textstyle{\frac{1}{4}}}
\newcommand{\cP}{\ensuremath{\mathcal{P}}}

\newcommand{\cPT}{\ensuremath{\mathcal{PT}}}

\newcommand{\bea}{\begin{eqnarray}}
\newcommand{\eea}{\end{eqnarray}}
\newcommand{\be}{\begin{equation}}
\newcommand{\ee}{\end{equation}}
\newcommand{\bal}{\begin{align}}
\newcommand{\eal}{\end{align}}

\usepackage{tikz} 
\usetikzlibrary{shapes.misc}

\usepackage[errorstop]{feynmp}
\newcommand{\rom}[1]{\uppercase\expandafter{\romannumeral #1\relax}}

\begin{document}

\thispagestyle{empty}

\begin{flushright}
{\small KCL-PH-TH/2024-{\bf 42}}
\end{flushright}

\vspace{0.4cm}

\begin{center}
\Large\bf\boldmath
Phases of \cb quartic \cbl scalar theories and \cPT ~symmetry 
\end{center}

\vspace{-0.2cm}

\begin{center}
{ Leqian Chen$^\dag$\blfootnote{$^\dag$~leqian.1.chen@kcl.ac.uk} and Sarben Sarkar$^\ddag$\blfootnote{$^\ddag$~sarben.sarkar@kcl.ac.uk} \\
\vskip0.4cm
{\it Theoretical Particle Physics and Cosmology, King’s College London,\\ Strand, London WC2R 2LS, UK}\\
\vskip1.4cm}
\end{center}

\begin{abstract}
For quantum mechanical anharmonic oscillator-type Hamiltonians, it is shown that there is a relation between the energy eigenvalues of parity symmetric  and \cPT-symmetric phases for weak coupling. The possibility of such a relation was conjectured by Ai, Bender and Sarkar on examining the imaginary part of the ground state energy using path integrals. In the weak coupling limit, we show that the conjecture is true also for the real part of the ground state energy and of the  excited state energies. However, the conjecture is false for strong coupling. The analogous relation for partition functions in zero spacetime dimensions is valid for many cases. \cb However $O(N)$ symmetric  multi-component scalar fields, with $N>1$  and  a quartic interaction, do not satisfy the conjecture for zero and one dimensional spacetime. The possibility that the conjecture is valid, for a single component field theory in higher dimensional spacetimes, is discussed in a simplified model. \cbl
\end{abstract}

\newpage

\hrule
\tableofcontents
\vskip.85cm
\hrule

\section{Introduction }
\label{sec:Introduction}

The study of $\cPT$-symmetric quantum theory has its roots in  work by Dyson on the divergence of perturbation theory in quantum electrodynamics (QED) \cite{PhysRev.85.631}. Dyson considered the weak coupling series in powers of $\alpha$, the fine structure constant. The classical Coulomb force is proportional to $\alpha$. Hence if $\alpha$ changes sign, a repulsive  Coulomb force between like charges  becomes  an attractive one. There is a non-analytic change in the physical behaviour at $\alpha=0$, i.e. the stability of the vacuum changes. If  weak coupling  perturbation series for physical quantities is convergent, then it is also analytic at $\alpha=0$. From the non-analytic behaviour at $\alpha=0$  we conclude that the series is merely formal and has zero radius of convergence. This formal series is often derived using Feynman graphs, which show  a factorial increase in the number of graphs \cite{Bender:1969si} with the order in the series. Such an increase is generic and points to a non-convergence of such series for interacting quantum field theories in an arbitrary number $D$ spacetime dimensions (a feature disconnected from issues of renormalisation). 

The argument of Dyson remained unchallenged for many decades. More recently, in the context of the $D=1$ anharmonic oscillator model \footnote{The anharmonic oscillator model is one of the simplest nontrivial field theories.}, a loophole in the argument  was pointed out  by Bender and coworkers \cite{R1,PhysRevLett.130.250404,Bender:2007nj} through the use of $\cPT$ symmetry. The   Hamiltonian for the anharmonic oscillator is $H=p^2+x^2+gx^4$ . For positive coupling constant $g$ the potential $V(x)=x^2+gx^4$ leads to an infinite number of positive-energy (bound) states. However, if $g<0$, it is now realised \cite{Bender:2007nj} that the Hamiltonian $H$ defines {\it two different phases \footnote{The use of the word phase is borrowed from critical phenomena. A possible order parameter is any nonzero expectation value of odd powers of $x$. }} depending on whether the global symmetry of the boundary conditions is parity
symmetric ($\cP$ symmetric) or parity-time symmetric ($\cPT$-symmetric). These two phases are obtained by changing the sign of $g$ from positive to negative in \emph{two different} ways:

(i) One way  that the sign of $g$ can be changed is to deform the class of potentials $V(\theta)= x^2+ge^{i\theta}x^4~~(g>0)$ by varying $\theta$ smoothly from 0 to $\pm\pi$. From WKB analysis of the Schr\"odinger equation, on rotating to $\theta=\pi$, we obtain unstable states that \emph{decay} in time because there are outgoing waves at $\pm\infty$. Alternatively, on rotating to $\theta=-\pi$, we find states that \emph{grow} in time because there are incoming waves from $\pm\infty$. A parity reflection interchanges these two physical configurations.

(ii) A different way to change the sign of $g$  is to deform the class of potentials \cite{R1} $V_\pm (\delta)=x^2+gx^2(\pm ix)^\delta~~(g>0)$  by
smoothly varying $\delta $ from $0$ (harmonic oscillator) to $2$ (inverted quartic oscillator). With this deformation, it can be shown 
that $V_\pm(2)$ has a discrete positive-definite bound-state spectrum \cite{R1}. The bound states have two possible configurations,
incoming waves at $x=-$$\infty$ and outgoing waves at $x=+$$\infty$ or outgoing waves at $x=-$$\infty$ and incoming waves at
$x=+$$\infty$. These \emph{unidirectional-wave configurations} are separately invariant under $\cPT$ reflection \cite{R2}. The states and
corresponding energy levels are observed in the laboratory by using cold-atom scattering experiments \cite{PhysRevLett.130.250404}.
Thus, an upside-down quartic-potential Hamiltonian defines two possible phases, one having complex energies
and the other having real energies. These phases are distinguished by having different global symmetries, either
\cP~ invariance or \cPT~ invariance. The positive discrete energy levels of the right-side-up quartic potential and the
\cPT -symmetric upside-down quartic potential are not analytic continuations of one another and thus there really is a
discrete nonanalytic jump at $g = 0$; it is this discontinuity that underlies the divergence of perturbation theory for
the quartic anharmonic-oscillator model and can be understood using the theory of resurgence \cite{Delabaere:1999hea,Dorigoni:2014hea}. 

The case of the pure \cPT~anharmonic oscillator  has also been put on a rigorous footing in \cite{Dorey:2001uw}, using methods from the study of integrable systems. \cb This latter work focussed on quantum mechanics, but the use of complexified actions has implications for higher-dimensional quantum field theories.  However the rigour of the analysis cannot be maintained in higher dimensions since in \cite{Dorey:2001uw} \cPT-symmetric quantum mechanics  was connected to ODE/IM (Ordinary Differential Equation/Integrable Model) correspondence, which maps the spectral problem of specific ordinary differential equations to the energy spectrum in integrable quantum models. For example, the eigenvalues of a Schr\"odinger-like equation can correspond to the energy levels of models such as the KdV or sine-Gordon model. The generalisation to higher -dimensional theories remains to be sufficiently developed for applications to \cPT~ symmetry. This also applies to coupled ODEs which are related to higher rank Lie algebras (such as $so(n)$) or multi-component field theories, where each component can represent a different field or degree of freedom.

\cbl Recently the above issues were discussed within the context of a path integral formulation \cite{R3b,Croney:2023gwy}, \cb which is an intuitive but nonrigorous tool requiring numerical analysis \cbl. This formulation allows, in principle, a generalisation to field theory  at weak coupling, where steepest descent paths of $D=0$ are replaced by Lefschetz thimbles \cite{R32w,Aarts2014-nz} for $D \ge 1$, paths in function space.  One phase (or thimble configuration) is $\cP$-symmetric
and has an unstable vacuum state and the other is $\cPT$-symmetric and appears to have a stable vacuum state \footnote{Unlike the positive-coupling constant theory, in $D=4$ the $\cPT$ -symmetric negative-coupling-constant theory is asymptotically free  \cite{Symanzik1975-mi} } . Recently a conjectured connection between the two phases was proposed \cite{R3b} (see below). The conjecture was made partially to stimulate investigations into the phases involving \cPT ~symmetry and unravel deep connections with non-perturbative aspects of corresponding Hermitian theories. It is hoped that the generalisation to field theories with \cPT -symmetry will lead to radically new examples  of models beyond the Standard Model \cite{R3.1, R3.2, R3.3, R3.4, R3.5, R3.6, R3.7, R3.8, R3.9, R3.10, R3.11, R3.12, R3.13, R3.14, R3.15,Mavromatos:2024ozk,Chen:2024bya,Sablevice:2023odu,Jones-Smith:2009qeu}. 

 If  mathematical connections between phases can be established, then it would be possible to draw on the considerable number of results \cite{LeGuillou:1990nq} for Hermitian systems to make inferences about~ \cPT- symmetric theories .
The calculation in \cite{R3b}  is made in terms of partition functions. $\mathcal Z$ for a scalar field  theory with scalar $\phi$, corresponding to potentials $V_{\rm Herm}\left( \phi \right)  =\frac{1}{2} m^{2}\phi^{2} +\frac{1}{4} \lambda \phi^{4} $ 
and $V_{\cPT}\left( \phi \right)  =\frac{1}{2} m^{2}\phi^{2} -\frac{1}{4} g \phi^{4}  $ and $m \neq 0$. The conjecture is that

\be
\label{eq:conj}
\log \mathcal Z_{\cPT}\left( g\right)  =\frac{1}{2} \log \mathcal Z_{\rm{Herm}}\left( \lambda =-g+i0^{+}\right)  +\frac{1}{2} \log Z_{\rm{Herm}}\left( \lambda =-g-i0^{+}\right). 
\ee
In particular, for the \emph{ground state energy} $E_0$ (in $D=1$), this implies a related conjecture that 
\be
\label{eq:conj1}
E_{0,PT}\left( g\right)  =\frac{1}{2} E_{0,{\rm Herm}}\left( \lambda =-g+i0^{+}\right)  +\frac{1}{2} E_{0,{\rm Herm}}\left( \lambda =-g-i0^{+}\right).  \ee
It has been verified \cite{R3b}  only that the imaginary part of $E_0$ satsfies \eqref{eq:conj1}, but for the real part the relation is conjectural. This verification follows from the cut-plane analyticity of the ground state energy in the $g$ plane \cite{Bender1978-pk,Simon1991FiftyYO}, which is also conjectured to be valid for the partition function \cite{Figerou:1981xu}. From  arguments  in terms of path-integrals \cite{R3b,Figerou:1981xu} (which lead to \eqref{eq:conj})  we  expect that a \emph{similar} relation to \eqref{eq:conj1} applies for the \emph{excited} energy levels \footnote{The argument is less direct for excited energy levels. The path integral formulation in terms saddle points and 
Lefschetz thimbles \cite{R32w,Ai:2019fri,R32w}  holds also for the calculation higher order Greens functions. Greens functions contain all the information about the theory including the energies of excited states.}. \cb Such a relation is \emph{not} contained in the original conjecture \cite{R3b}, and should be regarded as a speculation.\cbl

\cb A conjecture of this type is made within the weak coupling framework for a single component quartic scalar theories \cite{R3b}; it is natural to consider whether the conjecture might also hold for theories with $O(N)$ symmetric scalar field theories\footnote{\cb There are many interesting \cPT ~symmetric theories such as \cite{Guo:2022jyk,Gasparian:2022acf}, which lie outside the framework of this conjecture .\cbl}.  Given the importance of continuous symmetries for physics, it is useful to discuss the conjecture within this context\footnote{A $N$-component theory  has a somewhat different semi-classical analysis since tunneling phenomena are controlled by most probable escape paths \cite{Weinberg:2012pjx}.}.\cbl The conjecture is motivated by a path integral formulation \cb and analysis based on   semiclassical methods, \cbl using \emph{ a Lefschetz thimble} \cite{R32w}  passing through both trivial and non-trivial fixed points\footnote{The language of fixed points is accurate for $D=0$ and ordinary integrals. The saddle points in $D=1$ and higher $D$ are functions (such as instantons).}. The behaviour near the trivial fixed point has an asymptotic \emph{weak coupling} expansion. It is assumed that instanton solutions can be found. The conjecture should not be expected to hold if any of these assumptions are not valid~\cite{Lawrence:2023woz}. It is not known how to go to higher orders in semi-classical analysis \cite{R15} using WKB  or resurgence methods~\cite{Kamata:2023opn} for models which are not  single component scalar theories for $D \ge 1$. 

At one level, \cb for our class of theories \cbl, Hermitian and \cPT-symmetric Hamiltonians are ``trivially'' related
by having different signs in front of couplings. However the effect of these signs is nontrivial. We shall consider in detail the conjecture for various potentials in $D=1$, both with zero and nonzero $m$, using the Schr\"odinger equation, which is more versatile than the path integral for $D=1$. The case of zero $m$ is non-perturbative and the conjecture, based on weak coupling, may a priori fail. The path integral formulation is natural in $D=0$, and because of its simplicity it is possible to investigate  cases with \emph{multi-component} fields \cb with a continuous $O(N)$ symmetry, without the need for a semiclassical analysis, by using hyperspherical coordinates. A weak coupling analysis using steepest descent paths, with \cPT ~symmetry, is not known for general $N$ \cite{Berry1991-me,Howls1997HyperasymptoticsFM} \cbl.  Our aim here is to give a detailed examination of the conjecture in $D=0,1$ for models with the following potentials where $g>0$:
\begin{align}
\label{potentials}
  V_{1,2}\left( x\right)  &=\frac{1}{2} m^{2}x^{2}\pm \frac{1}{4} gx^{4},\nonumber\\
   V_{3, 4 }\left( x\right)  &=\pm \frac{1}{4} gx^{4} ,\nonumber\\
    V_{5, 6 }\left( x\right)  &=-\frac{1}{2} m^{2}x^{2}\pm \frac{1}{4} gx^{4}.
   \end{align}
and $V_{2}\left( x\right),V_{4}\left( x\right) $ and $V_{6}\left( x\right)$ are associated with potentials, which have a negative quartic coupling \cb and $N$-component versions of these models \cbl.
The analysis that we perform relies on numerical solutions of the Schr\"odinger equation \cite{Bender1993-hk}, WKB analysis and Borel resummation of  perturbation theory \cite{Dunne:2015eaa,LeGuillou:1990nq,R1}. It is however instructive to first consider the partition function for the $D=0$ case of the above potentials first, which is expressed through ordinary integrals and so can be investigated analytically, \cb without the need to resort to semiclassical analysis. The $D=1$ case  is nontrivial and will be considered with both strong and weak coupling using numerical methods; the analysis is also extended to the  $N$ component case \cite{Ogilvie2008-yj} which we show on symmetry grounds to be incompatible with the conjecture. We shall discuss the  $D>1$ case within the context of a model of Collins and Soper \cite{Collins:1977dw} where the nonperturbative fixed points are more straightforward to study than in a conventional quartic scalar field theory.   \cbl

\section{$D=0$ case}
The toy $D=0$ model has been studied earlier in~\cite{R2,Bender:2012loh} and is a simple (but still instructive) example of a trivial field theory \footnote{Spacetime has just one point.}. We first consider the integral for the Hermitian partition
function
\begin{align}
\label{ee3}
\widetilde{Z}_{\rm Herm}(\lambda)&=\int_{-\infty}^\infty
{\rm d}x\,\exp\left(-V(x) \right),
\end{align}
where $V(x)$ denotes one of the potentials given in \eqref{potentials}. For $V_{2,4,6}$ the integral is not convergent. Let us consider $V_{2}$ for definiteness and associate with it a $\cPT$ variant $V_{2}(\delta)$ which leads to 

\begin{equation}
\label{e4}
\widetilde{Z}^{\cPT}(g)=\lim_{\delta\to 2}\int_{C_{\cPT}}dx\,
\exp\left(-\half m^2x^2-\fourth g x^2\left(i x\right)^\delta\right),
\end{equation}
where $C_{\cPT}$ is a continuous contour that terminates in the \cPT-symmetric
Stokes wedges $-\frac{3}{8}\pi<\arg(x)<-\frac{1}{8}\pi$ and $-\frac{7}{8}\pi<
\arg(x) <-\frac{5}{8}\pi$ \cite{Bender:2007nj}. The contour $C_{\cPT}$ is not unique; deforming a
contour to another, which also terminates in the \cPT-symmetric Stokes wedges, leaves the value of the
integral unchanged due to Cauchy's theorem. It is convenient to make a change of variables:
$\varphi =mx,\  \epsilon =\frac{g}{m^{4}}$ and $t=\varphi \sqrt{\epsilon}$. This leads to 
\be
\mathcal{Z}_{1}(\epsilon)=\frac{1}{m\sqrt{\epsilon}} \int_{-\infty}^{\infty} dt\exp \left[ -\frac{1}{\epsilon} \left( \frac{t^{4}}{4} +\frac{t^{2}}{2} \right) \right]
\ee
in the Hermitian case and, for the $\cPT$-symmetric case $\mathcal{Z}_{2}$, 
\be
\mathcal{Z}_{2}(\epsilon)=\int_{C_{PT}} dt\  \frac{1}{m\sqrt{\epsilon}} \exp \left( \frac{1}{\epsilon} \left( \frac{t^{4}}{4} +\frac{t^{2}}{2} \right) \right),
\ee
defined using the complex $C_{\cPT}$ contour in the $t$-plane. 
The saddle points (in the scaled variables) are not dependent on the parameters. A possible $C_{\cPT}$ contour is $$\left\{ t=r\exp \left( i\theta \right) |\left[ \theta =-\frac{\pi}{4} ,0<r<\infty \right] \cup \left[ \theta =-\frac{3\pi}{4} ,0<r<\infty \right] \right\}.$$ 
It is straightforward \footnote{The integrals are evaluated in terms of modified Bessel functions. We use that for modified Bessel functions of the second kind and $\nu$ noninteger $K_{\nu}\left( x \right) =\frac{\pi}{2} \frac{I_{-\nu}\left( x \right) -I_{\nu}\left( x \right)}{\sin \left( \nu \pi \right)}$.} to show that~\cite{R3b}
\be
\label{nonanal}
\frac{1}{2} \left[ \mathcal{Z}_{1}\left( -\epsilon +i0+ \right) +\mathcal{Z}_{1}\left( -\epsilon -i0+ \right) \right] =Z_{2}\left( \epsilon \right),
\ee
which is a non-analytic relation between a Hermitian and a \cPT-symmetric theory.
This example was suggestive that relations such as  \eqref{eq:conj1} hold.  The potentials in \eqref{potentials} differ by having different signs for the $m$-term and $g$-term.

  For the case  $V_6$, we need to consider the related integral
\be
\mathcal{Z}_{6}(\epsilon)=\int_{C_{PT}} dt\  \frac{1}{m\sqrt{\epsilon}} \exp \left( \frac{1}{\epsilon} \left( \frac{t^{4}}{4} +\frac{t^{2}}{2} \right) \right).
\ee
On evaluating this integral, we get $\mathcal{Z}_{6}(\epsilon)=\frac{1}{2m\sqrt{\epsilon}} \exp \left( -\frac{1}{8\epsilon} \right) K_{-1/4}\left( \frac{1}{8\epsilon} \right)$. Similarly 
we consider $\mathcal{Z}_{5}(\epsilon)$, which is given by
\be
\mathcal{Z}_{5}\left( \epsilon \right) =\int_{-\infty}^{\infty} \frac{dt}{m\sqrt{\epsilon}} \exp \left( \frac{1}{\epsilon} \left( -\frac{t^{4}}{4} +\frac{t^{2}}{2} \right) \right),
\ee
and can be evaluated to give
$Z_{5}\left( \epsilon \right) =\frac{\pi \exp \left( \frac{1}{8\epsilon} \right)}{2m\sqrt{\epsilon}} \left( I_{-\frac{1}{4}}\left( \frac{1}{8\epsilon} \right) +I_{\frac{1}{4}}\left( \frac{1}{8\epsilon} \right) \right)$. It is straightforward to show that a relation such as \eqref{nonanal} does also hold for the pair $\mathcal{Z}_{5}$ and $\mathcal{Z}_{6}$.
 \vskip .1cm
 For $D=0$ it is possible to examine cases to see how the conjecture may break down \cb when multicomponent fields are considered \cbl. We   consider another strong coupling case  (with $m=0$) given by
\be
\mathcal{Z}_{7}(g)=\int_{-\infty}^{\infty} dx \, x^{2}\exp \left( -\frac{g}{4} x^{4} \right),
\ee
which is related to the $O(3)$ invariant theory\footnote{This case readily generalises to an odd number of oscillators.} integral
\be
\int_{-\infty}^{\infty} dz\int_{-\infty}^{\infty} dy\int_{-\infty}^{\infty} dx\  \exp \left( -\frac{g}{4} \left( x^{2}+y^{2}+z^{2} \right)^{2} \right),
\ee
with $g>0$. We make the ansatz that the \cPT-symmetric version is obtained by allowing $g \to -g$ and by choosing the $\cPT$ contour in the complex $x$ plane. \cb This bypasses the issue of choosing \cPT -symmetric contours for the $x$, $y$ and $z$ variables explicitly in a consistent manner; this is possible because there is no need for a semiclassical analysis in $D=0$.\cbl

\subsection{The strong coupling case}

We will first consider $V_{3}(x)$ and $V_{4}(x)$. It is easy to show that
$$\mathcal{Z}_{3}\left( g \right) =\frac{\Gamma \left( \frac{1}{4} \right)}{\sqrt{2} g^{1/4}}$$ and
$$\mathcal{Z}_{4}
\left( g \right) =\frac{\Gamma \left( \frac{1}{4} \right)}{2g^{1/4}}.$$
Moreover,
\be
\mathcal{Z}_{3}\left( -g+i0^{+} \right) +\mathcal{Z}_{3}\left( -g-i0^{+} \right) =\frac{\Gamma \left( \frac{1}{4} \right)}{g^{1/4}}.
\ee
Consequently
\be
\mathcal{Z}_{4}\left( g \right) =\frac{1}{2} \left[ \mathcal{Z}_{3}\left( -g+i0^{+} \right) +Z_{3}\left( -g-i0^{+} \right) \right],
\ee
which supports the findings in \cite{R3b}.

Turning to the case $\mathcal{Z}_{7}$ we find 
\be
\mathcal{Z}_{7}\left( g \right) =\frac{\sqrt{2}}{g^{3/4}} \Gamma \left( 3/4 \right).
\ee
The $\cPT$ version is 
\be
\begin{gathered}\mathcal{Z}_{8}\left( g \right) =\int_{C_{PT}} dx\  x^{2}\exp \left( \frac{g}{4} x^{4} \right)=-\frac{\Gamma \left( 3/4 \right)}{g^{3/4}} \  \end{gathered}.
\ee
We find 
\be
\mathcal{Z}_{8}\left( g \right) =\frac{1}{2} \left[ \mathcal{Z}_{7}\left( -g+i0^{+} \right) +Z_{7}\left( -g-i0^{+} \right) \right],
\ee
which is similar to the result in \cite{R3b}.
\cb
\subsection{The multi-component case}
For an $O(M)$ symmetric theory with $M=2N+1$ (where $N$ is a positive integer) we consider 
\be
\begin{gathered}\mathcal{Z}_{2N+5}\left( g \right) =\int_{-\infty}^{\infty} dx\  x^{2N}\exp \left( -\frac{m^2}{2}x^{2}-\frac{g}{4} x^{4} \right) \end{gathered}
\ee
and
\be
\begin{gathered}\mathcal{Z}_{2N+6}\left( g \right) =\int_{C_{PT}} dx\  x^{2N}\exp \left( -\frac{m^2}{2}x^{2}+ \frac{g}{4} x^{4} \right).  \end{gathered}
\ee
It is straightforward to show that
\begin{multline}
\mathcal{Z}_{2N+5}\left( g \right) =2^{N-\frac{1}{2}}
   g^{-\frac{N}{2}-\frac{3}{4}}
   ( \sqrt{g} \Gamma
   \left(\frac{N}{2}+\frac{1}{4}\right) \,
   _1F_1\left(\frac{N}{2}+\frac{1}{
   4};\frac{1}{2};\frac{m^4}{4
   g}\right)\\
   -m^2 \Gamma
   \left(\frac{N}{2}+\frac{3}{4}\right) \,
   _1F_1\left(\frac{N}{2}+\frac{3}{
   4};\frac{3}{2};\frac{m^4}{4
   g}\right) )
   \end{multline}
and
\begin{multline}
\mathcal{Z}_{2N+6}\left( g \right) =2^{N-\frac{1}{2}}
   g^{-\frac{N}{2}-\frac{3}{4}}
   ( \sqrt{g} \Gamma
   \left(\frac{N}{2}+\frac{1}{4}\right) \,
   _1F_1\left(\frac{N}{2}+\frac{1}{
   4};\frac{1}{2};\frac{-m^4}{4
   g}\right)\cos \left( \frac{\pi}{2} (N+\frac{1}{2} ) \right)\\
   +m^2 \Gamma
   \left(\frac{N}{2}+\frac{3}{4}\right) \,
   _1F_1\left(\frac{N}{2}+\frac{3}{
   4};\frac{3}{2};\frac{-m^4}{4
   g}\right)\cos \left( \frac{\pi}{2} (N-\frac{1}{2} ) \right) )
\end{multline}
(We note that $\cos \left( \frac{\pi}{2} (N+\frac{1}{2} ) \right) =\frac{(-1)^{n}}{\sqrt{2}} \left( \delta_{N\  2n} -\delta_{N\  2n+1} \right)$ and $\cos \left( \frac{\pi}{2} (N-\frac{1}{2} ) \right) =\frac{(-1)^{n}}{\sqrt{2}} \left( \delta_{N\  2n} +\delta_{N\  2n+1} \right)$ where $n$ is a positive integer; ${}_{1}F_{1}(a;b;z)$ is the Kummer confluent hypergeometric function, which satisfies ${}_{1}F_{1}(a;b;z)=\exp \left( z \right) {}_{1} F_{1}(b-a;b;-z)$ and ${}_{1}F_{1}(a;b;0)=1$ for $b \neq -n$.)
\vskip .3cm
\noindent Hence, for arbitrary $N$ and $g$ we see clearly that
\be
\mathcal{Z}_{2N+6}(g)\neq \frac{1}{2} \left[ \mathcal{Z}_{2N+5}\left( -g+i0^{+} \right) +\mathcal{Z}_{2N+5}\left( -g-i0^{+} \right) \right],
\ee
and this is indicative of a failure of the conjecture in \cite{R3b}, certainly in its simplest form, and for \emph{arbitrary} coupling (i.e. arbitrary $m$). Multicomponent scalar theories provide, more generally,  examples \cite{Lawrence:2023woz} where the conjecture is invalid for $D>0$ \footnote{In a path integral formulation this is related  to the difficulty of finding steepest descent paths simultaneously in all components of the field, while respecting \cPT~ symmetry. }. The failure of the conjecture, for the multicomponent case,  is demonstrated for $D=1$ in Sec.{(\ref{Nanharmonic})}. 
\cbl

\section{$D=1$ case}
Although $D=0$ provides simple examples \cite{R3b}, which can be analysed fully, physical applications occur for $D \ge 1$. The $D=1$ case represents quantum mechanics, which is a   field theory (albeit simplified, with no particle scattering for example) \cite{Bender:1969si,Bender:1973rz}.The analyticity properties of the energy eigenvalues in the complex coupling plane are studied in detail by Bender and Wu \cite{Bender:1969si,Bender:1973rz} in an important work. In \cite{R3b} we return to the model in its  \cPT ~symmetric formulation \cite{R1} and make a semiclassical (weak coupling \cite{Ivanov_1998}) \emph{path-integral} calculation for $m \neq 0$; the conjectured relation is shown to hold for the imaginary part of the ground state energy. In \cPT~ symmetric theories the imaginary of the ground state energy is zero and so the result in \cite{R3b} needs to be extended to real parts, not just for the ground state but also for (low-lying) excited states. 

For the ground state we can use the path integral formulation \cite{Zinn-Justin2004-wk} to calculate the real part of the ground state energy for the Hermitian anharmonic oscillator. We shall illustrate some of the features of this (divergent) series when used for analytic continuation in the coupling (first noted in \cite{Bender:1969si}). The path integral method is not convenient for investigating excited states, whereas  the Schr\"{o}dinger  equation \cite{Muller-Kirsten2006-jq} allows an investigation of the conjecture in both the weak 
and strong coupling \cite{Ivanov_1998} regimes for low energy eigenstates and their eigenvalues, which can then be compared with eigenvalues obtained from the Schr\"{o}dinger equation for \cPT ~quantum mechanics.The methods in \cite{Bender1993-hk} are used to study the latter.

\subsection{Path integrals and the weakly coupled anharmonic oscillator ground state}

The path integral formulation \cite{Feynman2010-ew} is natural for investigating the ground state energy of a quantum mechanical system. Furthermore it opens up the possibility of semiclassical analyses in higher dimensional spacetime \cite{R3b} (involving Lefschetz thimbles \cite{R32w}) of quantities of physical importance. For a Hamiltonian, 
$$H=\frac{1}{2} p^{2}+V\left( x \right),$$ we consider the partition function $Z\left( \beta \right) =Tr\  e^{-\beta H}$. The ground state energy $E_0$ is given by
$$E_{0}=-\lim_{\beta \rightarrow \infty} \log Z\left( \beta \right).$$
It is known that $Z\left( \beta \right)$ has a path integral representation \cite{Zinn-Justin2004-wk}:
$$Z\left( \beta \right) =\int Dx\  \exp (-S[x]),$$
where
$$S[x]=\int_{-\beta /2}^{\beta /2} dt\left[ \frac{1}{2} \left( \dot{x} \left( t \right) \right)^{2} +V\left( x\left( t \right) \right) \right]$$
and $x\left( -\frac{\beta}{2} \right) =x\left( \frac{\beta}{2} \right)$.
We consider
$V\left( x \right) =\frac{1}{2} x^{2}+\frac{g}{4} x^{4}$. $Z\left( \beta \right)$ can be expanded as a series in $g$. In the limit $\beta \to \infty$ we obtain a conventional (zero-temperature) field theory with propagator 
$$D\left( \tau ,\tau^{\prime} \right) =\frac{\exp (-\left| \tau -\tau^{\prime} \right| )}{2}$$
and an interaction $4$-vertex with coupling $\frac{g}{4}$. This limit also prevents the direct investigation of the excited states. Only connected diagrams contribute to $\log Z\left( \beta \right)$ and we have a series 
$$E_{0}\left( g \right) =\frac{1}{2} +\sum_{k=1}^{\infty} a_{k}\left( \frac{g}{4} \right)^{k}.$$
These coefficients were first calculated to high order by Bender and Wu \cite{Bender:1969si,Bender:1973rz} and more recently in \cite{Sulejmanpasic:2016fwr}. The lowest three are: $a_{1}=\frac{3}{4} ,\  \  a_{2}=-\frac{21}{8} ,\  a_{3}=\frac{333}{16}.$ However it is well known that this is a divergent series due to the singularity structure of the full function $E_{0}\left( g \right)$ in the complex $g$-plane and so cannot be used directly to investigate the conjectured connection between \cPT-symmetric energy eigenvalues and those of the corresponding Hermitian theory. Indeed using WKB methods and matched asymptotics \cite{Bender:1969si,Bender:1968sa} Bender and Wu showed (see also \cite{Simon:1970mc}) that the singularity at $g=0$ is not isolated. The energy eigenvalues can be continued in the complex $g$ plane, but there are an infinite number of branch point singularities in the complex $g$ plane, which accumulate at $g=0$ \cite{Simon:1970mc}. One such branch cut is along the negative $g$-axis.

\subsection{Massive anharmonic oscillator}

Because of the difficulties of continuing directly in the coupling $g$, we shall follow a method given in \cite{Liang:2004na,Muller-Kirsten2006-jq}, which is a modified WKB analysis and seems to be little known. It is related to the analysis found in \cite{Bender:1969si,Bender:1973rz}, but considers an expansion (related to a weak coupling expansion) in inverse powers of the strength of the harmonic term. In order to compare with the notation in \cite{Muller-Kirsten2006-jq}, we shall take the negative coupling  $g=-2c^{2}\pm i0^{+}$ and $m=h^{2}$ where $c$ and $h$ are real. 

In this notation, the Schr\"odinger equation is
\begin{equation}
\label{negSch}
    \left[-\frac{\mathrm{d}^2}{\mathrm{~d} x^2}+V(x)\right] \psi(x)= E \psi(x),
\end{equation}
with the potential
\begin{equation}
\label{muller}
    V (x) = \frac{1}{4} h^4 x^2 - \frac{1}{2} c^2 x^4.  
\end{equation}
Let us write $w=hx$ and parametrise $E$ as $E=\frac{1}{2} qh^{2}+\frac{\Delta \left( q,h \right)}{2h^{4}}$. Then \eqref{negSch} becomes
\be
\mathcal{D}_{q}\left( w \right) \psi \left( w \right) =-\frac{1}{h^{6}} \left( \Delta \left( q,h \right) +c^{2}w^{4} \right) \psi \left( w \right)
\ee
with $\mathcal{D} _{q}\left( w \right) \equiv 2\frac{d^{2}}{dw^{2}} +q-\frac{w^{2}}{2}.$ In this method there is an expansion in inverse powers of $h$, which is reminiscent of conventional strong coupling expansions. The shape of $V(x)$ has associated with it natural domains in $x$. The positions of the maxima of $V(x)$, $x_{\pm}$,  are determined from $V^{\prime}(x_{\pm})=0$ and so $V^{\prime\prime}(x_{\pm})=h^{4}$ and $V(x_{\pm})=\frac{h^{8}}{2^{5}c^{2}}$. Solutions valid near $x=0$ are obtained from 
\be
\mathcal{D}_{q}\left( w \right) \psi \left( w \right) =0,
\ee
whose solution is the parabolic cylinder functions $D_{\frac{1}{2} \left( q-1 \right)}\left( w \right)$ and is valid upto $x\simeq O\left( \frac{1}{h^{2}} \right)$. Away from this interval around  $x=0$ in \cite{Muller-Kirsten2006-jq} , the parabolic cylinder solution is matched to a WKB-type solution 
\be
\psi \left( x \right) =A\left( x \right) \exp \left[ \pm i\int_{}^{x} dx\left\{ -\frac{h^{4}x^{2}}{4} +\frac{c^{2}x^{4}}{2} \right\}^{1/2} \right].
\ee
After some intricate matching of asymptotic expansions, it is shown in \cite{Muller-Kirsten2006-jq}
that the energy eigenvalues are 
\begin{align}
\label{HermMassiveEnergy}
    {E}_{q_0}=&\frac{q_0}{2} h^2-\frac{3 c^2}{4 h^4}\left(q_0^2+1\right)-\frac{q_0 c^4}{h^{10}}\left(4 q_0^2+29\right)+\mathcal{O}\left(\frac{1}{h^{16}}\right) \nonumber\\
    &\pm \mathrm{i} \frac{2^{q_0} h^2\left(\frac{h^6}{2 c^2}\right)^{q_0 / 2}}{(2 \pi)^{1 / 2}\left[\left(q_0-1\right) / 2\right] !} \mathrm{e}^{-h^6 / 6 c^2}
\end{align}
where $q_0 = 2k +1$ and $k$ labels the energy level. The imaginary part arises from the divergence of the perturbation theory, expressed in terms of an instanton; a cut, along the negative $g$ axis in the complex $g$ plane for the energies, leads to a discontinuity in the imaginary part expressed with $\pm$ in \eqref{HermMassiveEnergy} \cite{Bender:1969si,Bender:1973rz} . These expressions represent the parity symmetric phase of the anharmonic oscillator with negative coupling.
Since we  recover $V_2(x)$ by making the substitution $h^2 \rightarrow \sqrt{2} m$ and $c^2 \rightarrow \frac{1}{2} g$, the real part of energy eigenvalues become

\begin{equation}
\label{energy expression muller}
    E_k (m,g) \approx \frac{(2
   k+1) m}{\sqrt{2}}
   -\frac{3 g \left(2 k^2+2 k+1\right)}{8 m^2}
   -\frac{g^2 \left(32 k^3+48 k^2+82 k+33\right)}{16 \sqrt{2} m^5}
   +\mathcal{O}\left(\frac{1}{m^{8}}\right). 
\end{equation}
 On choosing $g=4$,  the energy spectrum varies with the value of $m$ as given below.

\begin{table}[htbp]
    \centering
    \begin{tabular}{>{\centering\arraybackslash}p{0.15\linewidth}>{\centering\arraybackslash}p{0.15\linewidth}>{\centering\arraybackslash}p{0.15\linewidth}>{\centering\arraybackslash}p{0.15\linewidth}>{\centering\arraybackslash}p{0.15\linewidth}>{\centering\arraybackslash}p{0.15\linewidth}}
         &  $m=\sqrt{2}$&  $m=\sqrt{10}$&  $m=\sqrt{20}$&  $m=\sqrt{200}$&  $m=\sqrt{2000}$\\ \hline
         $E_0$
&  -3.875&  2.01228&  3.07423&  9.99246&  31.622\\
         $E_2$&  -85.375&  7.78808&  14.5814&  49.9017&  158.104\\
         $E_4$&  -418.875&  6.87062&  24.1297&  89.6885&  284.574\\
         $E_6$&  -1196.38&  -4.17468&  31.1118&  129.351&  411.032\\
    \end{tabular}
    \caption{Energy spectrum of massive anharmonic oscillator at $g=4$ using \cbl the method of \cbl M\"uller-Kirsten \cite{Muller-Kirsten2006-jq}}
    \label{Muller-Kirsten energy spectrum 2}
\end{table}
\noindent The negative energy eigenvalues found for $m=\sqrt{2}$ are nonphysical. They are due  to the neglect of terms of higher order  $\mathcal{O}\left(\frac{1}{m^{8}}\right)$, which become significant when $m$ is small. The series is asymptotic rather than convergent and more terms will require an optimal truncation analysis. However we find, as $m$ gets larger, that our approximation  becomes  more accurate. Equation \eqref{HermMassiveEnergy} is an example of a transseries \cite{Dorigoni:2014hea} and has higher order subdominant multi-instanton contributions \cite{Kamata:2023opn}, which, even though they may have a real part, are exponentially suppressed in the weak coupling limit.

We can also compute the energy spectrum of the same differential equation \emph{numerically} in the \cPT -symmetric phase:$-\frac{1}{4}g x^4 =\frac{1}{4} g x^2 (i x)^\delta$ and let $\delta \rightarrow 2$. To do so, we deform the contour into the complex plane (see appendix \ref{RungeKutta}) inside a Stoke's wedge \cite{Bender:2007nj} and use a modified Runge–Kutta algorithm to calculate the energy eigenvalues. The energy spectrum is summarised in table \ref{massive numerical energy spectrum}
\begin{table}[htbp]
    \centering
    \begin{tabular}{>{\centering\arraybackslash}p{0.15\linewidth}>{\centering\arraybackslash}p{0.15\linewidth}>{\centering\arraybackslash}p{0.15\linewidth}>{\centering\arraybackslash}p{0.15\linewidth}>{\centering\arraybackslash}p{0.15\linewidth}>{\centering\arraybackslash}p{0.15\linewidth}}
         &  $m=\sqrt{2}$&  $m=\sqrt{10}$&  $m=\sqrt{20}$&  $m=\sqrt{200}$&  $m=\sqrt{2000}$\\ \hline
         $E_0$
&  1.15224&  2.04440&  3.08248&  9.99249&  31.6220\\
         $E_2$&  5.09893&  5.64812&  14.6620&  49.9021&  158.10413\\
         $E_4$&  10.4406&   9.14652&  24.0206&  89.6904&  284.57423\\
         $E_6$&  16.7016&  13.7260&  32.9800&  129.356&  411.03232\\
    \end{tabular}
    \caption{Energy spectrum of massive anharmonic oscillator at $g=4$ calculated numerically using the \cPT ~contour}
    \label{massive numerical energy spectrum}
\end{table}

One can compare table \ref{massive numerical energy spectrum} with table \ref{Muller-Kirsten energy spectrum 2}; it is evident that the values of the energy agree better in the limit of large $m$  compared to small $m$, and are better for the ground state compared to the excited states. This may be due to contribution of the higher order terms neglected in equation \eqref{energy expression muller}. It is worth noting that the method of M\"uller-Kirsten \cite{Muller-Kirsten2006-jq}  used for computing the real part of  energy eigenvalues for the potential $V_2(x)$,  agrees with the \cPT~ energy eigenvalues reasonably well \cbl in the limit of large $m$.\cbl

\subsection{Strong coupling case}

Although for $D=1$ the conjecture in \cite{R3b} is made within the context of weak coupling, it is of interest to consider its validity in strong coupling. 
The Schr\"{o}dinger  equation with potential $V_3$ has the Hamiltonian

\begin{equation}
\label{hamiltonian hermitian}
    \hat{H}(\hbar, g)=-\hbar^2 \frac{\mathrm{d}^2}{\mathrm{~d} x^2}+g x^4.
\end{equation}
When it acts on a wave function $\psi(x)$, it becomes a second order differential equation, which, with suitable boundary conditions, yields the energy spectrum. From a simple scaling argument due to Symanzik \cite{Simon:1970mc}, it is easy to see that there is no adjustable parameter. \cbl In fact, Symanzik shows that for this case, the energy, for different values of $g$,  scales as

\begin{equation}
E\left( g \right) =g^{1/3}E\left( 1 \right),
\end{equation}
where $E(1)$ is the energy eigenvalue for $g=1$.
\cbl
Consequently there is no weak coupling expansion. 
One can numerically calculate the value of the energy for different energy levels and different  $g$;  we summarise some of them in the table below (and note that the values satisfy Symanzik scaling, a useful check on the validity of the calculation).

\begin{table}[htbp]
    \centering
    \begin{tabular}{>{\centering\arraybackslash}p{0.25\linewidth}>{\centering\arraybackslash}p{0.25\linewidth}>{\centering\arraybackslash}p{0.25\linewidth}>{\centering\arraybackslash}p{0.25\linewidth}} 
         &  $g=1$&  $g=2$& $g=10$\\ \hline
         $E_0$&  1.06036&  1.33597& 2.28448\\ 
         $E_1$&  3.79967&  4.78729& 8.18615\\ 
         $E_2$&  7.45569&  9.39359& 16.0628\\ 
         $E_3$&  11.6447&  14.6715& 25.0878\\ 
         $E_4$&  16.2617&  20.4886& 35.035\\ 
    \end{tabular}
    \caption{Energy spectrum for the massless anharmonic oscillator (here we take $\hbar =1$)}
    \label{energy spectrum hermitian massless}
\end{table}
\noindent As a further check, we  compare some of the values in the table with the results obtained using WKB analysis by Voros and collaborators (see \cite{Albeverio1979-vt,AIHPA_1983__39_3_211_0}) and find  agreement. 

\cbl

We use Symanzik scaling to analytically continue the energy levels in the complex plane, i.e. $E_{n,{\rm Herm}}\left( -g+i0^{+}\right) =(-g+i0^{+})^\frac{1}{3}  E_{n,{\rm Herm}}\left( 1\right)$ and $E_{n,{\rm Herm}}\left( -g+i0^{-}\right) =(-g+i0^{-})^\frac{1}{3}  E_{n,{\rm Herm}}\left( 1\right)$. Denote $-g+i0^{+} = g e^{i (\pi - \epsilon)}$ and $-g+i0^{-} = g e^{-i (\pi - \epsilon)}$ where $\epsilon>0$ is infinitesimal and consider

\begin{equation}
\label{eq.conj30}
    \frac{1}{2} E_{n,{\rm Herm}}(-g+i 0^+) +     \frac{1}{2} E_{n,{\rm Herm}}(-g+i 0^-) =     \frac{1}{2} g^\frac{1}{3} E_{n,{\rm Herm}}(1) \equiv  \frac{1}{2} E_{n,{\rm Herm}}(g),
\end{equation}
which is the right hand side of equation \eqref{eq:conj1}. 

Now let us turn to the \cPT-symmetric  case, which is described by the Hamiltonian with the inverted potential $V_4(x)$. Again, we use numerical methods to calculate the energy spectrum \footnote{See the Appendix}. The energy spectrum for the Hamiltonian $H=-\frac{\hbar^2}{2m}\frac{\mathrm{d}^2}{\mathrm{~d} x^2}+g x^2(i x)^\delta$ at $\delta=2$ and $g=1,2$ and $10$ is given below:
\begin{table}[h]
    \centering
    \begin{tabular}{>{\centering\arraybackslash}p{0.25\linewidth}>{\centering\arraybackslash}p{0.25\linewidth}>{\centering\arraybackslash}p{0.25\linewidth}>{\centering\arraybackslash}p{0.25\linewidth}} 
         &  $g=1$&  $g=2$ &$g=10$\\ \hline
         $E_0$&  1.47714&  1.86109 &3.18242\\ 
         $E_1$&  6.00338&  7.56379 &12.93390\\ 
         $E_2$&  11.80243&  14.87013 &25.42757\\ 
         $E_3$&  18.45881&  23.25665 &39.76832\\ 
         $E_4$&  25.79179&  32.49562 &55.56673\\ 
    \end{tabular}
    \caption{Energy spectrum for the \cPT~massless anharmonic oscillator with inverted potential (here we take $\hbar =1,m=\frac{1}{2}$)}
    \label{energy spectrum PT massless}
\end{table}

One can compare the results in Table \ref{energy spectrum PT massless}  (which satisfy Symanzik scaling) with the values from Table \ref{energy spectrum hermitian massless}. \cbl It is easy to see that for every energy level, the value we obtained from the \cPT ~case  is always greater than the conjectured value from the Hermitian case. Indeed, from equation \eqref{eq.conj30} we know that the right hand side of the conjecture is half of the Hermitian energy.  Therefore, it is clear that \eqref{eq:conj1} for strong coupling is invalid in the $D=1$ case. \cbl

Also it has been proved (using both functional and canonical methods) that the energy spectrum for the \cPT~ massless anharmonic oscillator is identical to the energy spectrum (but not wavefunctions) of the following Hamiltonian with an anomaly term \cite{PhysRevD.74.025016,Jones:2006et} 
\begin{equation}
\label{hamiltonian hermitian anomaly}
    \tilde{H}( g)=-\frac{\hbar^2}{2m} \frac{\mathrm{d}^2}{\mathrm{~d} x^2}+4g x^4 -\hbar \sqrt{\frac{2g}{m}}x
\end{equation}
The energy spectrum below is calculated numerically with this Hamiltonian. The results in \\ Table \ref{energy spectrum PT massless} agree with the values from Table \ref{energy spectrum hermitian massless anomaly} to a good precision. Therefore, this is further proof that  \eqref{hamiltonian hermitian anomaly}  represents a spectrally equivalent Hamiltonian, which generates the \cPT~ energy spectrum.
\begin{table}[htbp]
    \centering
    \begin{tabular}{>{\centering\arraybackslash}p{0.25\linewidth}>{\centering\arraybackslash}p{0.25\linewidth}>{\centering\arraybackslash}p{0.25\linewidth}>{\centering\arraybackslash}p{0.25\linewidth}} 
         &  $g=1$&  $g=2$ &$g=10$\\ \hline
         $E_0$&  1.47714&  1.86109 &3.18242\\ 
         $E_1$&  6.00338&  7.56379 &12.9339\\ 
         $E_2$&  11.8024&  14.8701&25.4276\\ 
         $E_3$&  18.4588&  23.2567&39.7683\\ 
         $E_4$&  25.7918&  32.4956&55.5667\\ 
    \end{tabular}
    \caption{Energy spectrum for the Hermitian massless anharmonic oscillator with an anomaly term (here we take $\hbar =1,m=\frac{1}{2}$)}
    \label{energy spectrum hermitian massless anomaly}
\end{table}

\cbl  It is easy to see that, for strong coupling, the spectra of the Hermitian and \cPT-symmetric Hamiltonians are not related. Hence, we can conclude that there is no value of $g$ \cbl such that relation \eqref{eq:conj1} is valid and the Ai-Bender-Sarkar conjecture~\cite {R3b} does not hold for strong coupling (see WKB analysis in \cite{Kamata:2023opn} for an alternative argument using alien calculus). 
\cb
\subsection{The $N$-component anharmonic oscillator}
\label{Nanharmonic}
We  consider the Euclidean $O(N)$ symmetric model with \cPT -symmetric Lagrangian:
\be
L=\sum_{j=1}^{N} \left[ \frac{1}{2} \left( \partial_{t} x_{j} \right)^{2} +\frac{1}{2} m^{2}x_{j}^{2}-\lambda x_{j}^{4} \right] -\frac{g}{N} \left( \sum_{j=1}^{N} x_{j}^{2} \right)^{2}.
\ee
with $g>0, \lambda >0 $\footnote{We will later take the limit $ \lambda  \to 0$.}.
It is convenient to rewrite \cite{Ogilvie2008-yj,Nishimura:2008yx} the quartic terms so that $L$ becomes: $$\mathcal{L}=\sum_{j=1}^{N} \left[ \frac{1}{2} \left( \partial_{t} x_{j} \right)^{2} +\frac{1}{2} m^{2}x_{j}^{2} \right] -\sum_{j,k=1}^{N} x_{j}^{2}\Lambda_{jk} x_{k}^{2},$$
where $\Lambda =\lambda I+gP$ $I$ is the $N$-dimensional unit matrix and $P$, a projection, is a $N \times N$  matrix  with each element $1/N$. The technique \cite{PhysRevD.74.025016,Jones:2006et} used when $N=1$, for obtaining a Hermitian  isospectral Hamiltonian, can be adapted here \cite{Ogilvie2008-yj,Nishimura:2008yx}. We restrict to the case of  $\Lambda$ with all positive eigenvalues (a natural choice for \cPT ~symmetry). The elementary method  that was applied to $N=1$, can be extended to $N>1$ by making the substitution $x_{j}\rightarrow -2i\sqrt{c_{j}+i\psi_{j}}$. On incorporating the Jacobian of the transformation, the path integral representation of the partition function \cite{Jones:2006et}, is
\be
Z=\int \prod_{j} \frac{\left[ d\psi_{j} \right]}{\sqrt{det\left( c_{j}+i\psi_{j} \right)}} \exp \left[ -\int dt\left[ \mathcal{L}-\sum_{j} \frac{1}{32} \left( \frac{1}{c_{j}+i\psi_{j}} \right) \right] \right].
\ee
Auxiliary fields $h_j$ are used to write the determinant term as an exponential: 
\be
\nonumber
\prod_{j} \frac{1}{det\left[ \sqrt{c_{j}+i\psi_{j}} \right]} =\int \prod_{j} \left[ dh_{j} \right] \exp \left\{ -\int dt\left[ \frac{1}{2} \left( c_{j}+i\psi_{j} \right) \left( h_{j}-\frac{i\psi_{j} +1/4}{c_{j}+ic_{j}} \right)^{2} \right] \right\}.
\ee
Integration over $\psi_{j}$ can readily be done. In \cite{Ogilvie2008-yj} the fields $\sigma$ and $\overrightarrow{\pi} $ (which is a vector with $N-1$ components) are introduced such that 
\be
\sigma =\frac{1}{\sqrt{N}} \sum_{j} h_{j},
\ee
and $\sigma^{2} +\overrightarrow{\pi}^{2} =\overrightarrow{h}^{2}$. On making the redefinition $h_{j}=\frac{1}{\sqrt{N}} \sigma +\bar{h}_{j}$ such that $\sum_{j} \bar{h}_{j} =0$ and rescaling $\sigma \rightarrow \sqrt{\left( g+\lambda \right) /\lambda} \sigma$, $\mathcal{L}$ becomes \cite{Ogilvie2008-yj}
\be
\mathcal{L}=\frac{1}{2} \left( \frac{d}{dt} \sigma \right)^{2} ++\frac{1}{2} \left( \frac{d}{dt} \overrightarrow{\pi}^{2} \right) -m^{2}\sigma^{2} +\frac{4g}{N} \sigma^{4} +\frac{16g}{N} \sigma^{2} \overrightarrow{\pi}^{2} -\sqrt{2gN} \sigma.
\ee
It is important to note in $\mathcal{L}$ that:
\begin{itemize}
    \item the $N=1$ limit is reproduced ( on disposing of $\overrightarrow{\pi}$).
    \item $m$ can be $0$ or nonzero.
    \item $g$ is not assumed to be small.
    \item the $O(N)$ symmetry of the \cPT -symmetric formulation (for $N>1$) is seen to be broken to $O(N-1)$ in the isospectral Hermitian context.
\end{itemize}
 Given this change in symmetry, there is no analytic deformation of the $N$-component Hermitian  theory to get the spectra of the \cPT-symmetric case. This approach, which uses an equivalent Hermitian Hamiltonian, is also used by Bender and Sarkar \cite{Bender:2013mpa} for the study of the double-scaling limit of the $O(N)$ symmetric anharmonic oscillator.
 %%%%%%%%%%%%%%%%%%%%%%%%%%%%%%%%%%%%%%%%%%%%
 \cbl

\section{\cb The $D>1$ case \cbl}
\cb In weak coupling field theory, energy levels (which form a continuum) are found on using perturbation theory  in Fock space \footnote{\cb In the absence of interactions the Hamiltonian is solvable representing free particles which are labelled by momenta. Energy levels of the free theory are given by the sum of the energies of the non-interacting particles. When a small interaction is introduced perturbation theory that gives an expansion in powers of the coupling $g$. Higher-order corrections involve complicated integrals over intermediate states, represented by Feynman diagrams. The process of renormalisation may be needed to remove infinities. \cbl}. Such a series is a formal series and is usually divergent because of proliferation of Feynman graphs and, for cases such as  $D=4$, there is the added complication of ultraviolet and infrared divergences in perturbation theory.  

For $D>1$, the perturbation theory for quantities such as the beta function \cite{Brezin:1976vw}  (because it is simpler  than the continuum of  energy levels) or the partition function itself may be more suited  for investigating the conjecture. Understanding in detail the semiclassical path integral for a \cPT -symmetric 
 quartic scalar theory is \emph{unsolved} in higher dimensions such as $D=4$. The conjecture \cite{R3b}  was made within the context of a weak coupling semiclassical path integral formulation. A higher dimensional generalisation requires  a rigorous understanding of the structure of the Lefschetz thimbles \cite{Tanizaki:2014xba,Behtash:2015loa,Aarts2014-nz,Cristoforetti:2013qaa,R32w}, which is missing. The current understanding is mainly numerical and approximate, which does not  lend itself to an anlalysis of the conjecture.

 There is a model by Collins and Soper (CS) \cite{Collins:1977dw},  which treats more simply the complexification  of the Lagrangian (necessary to investigate the conjecture). This model, applied to $D=1$, \emph{is} successful in understanding, within the context of field theory,  the asymptotic behaviour of the perturbation series of the ground state energy of the anharmonic oscillator (studied first by Bender and Wu \cite{PhysRevD.7.1620}). CS \cite{Collins:1977dw} obtain exact agreement with the quantum mechanical calculation in \cite{PhysRevD.7.1620}\footnote{\cb For $D>1$ we will see that the nonperturbative saddle points and their contributions, required in a weak coupling expansion of the partition function  lead to imaginary contributions which formally cancel in a \cPT -symmetric formulation, independent of any renormalisation procedures.\cbl} .

 The approach of CS \cite{Collins:1977dw}  is formally generalisable to higher dimensions since, for their functional, the steepest descent path is made complex through a single complex scalar variable  $\chi$. The Hamiltonian considered is 
 $$H=\frac{1}{2} \pi^{2} +\frac{m^{2}}{2} \phi^{2} +\frac{1}{2} \left( \nabla \phi \right)^{2}+\frac{g}{4} \phi^{4},$$
where $\pi$ is the canonical momentum for $\phi$ and $\nabla$ is the spatial derivative in $D-1$ dimensions. The partition function is defined as 
 \be
 Z\left( g,T \right) =Tr\left[ e^{-H(g)T} \right] /Tr\left[ e^{-H(0)T} \right]=\exp \left( -E\left( g,T \right) T \right) /\exp \left( -E\left( 0,T \right) T \right).
 \ee
where $T$ is Euclidean time. CS assume that $Z\left( g,T \right)$ has cut $g$-plane analyticity i.e. it is analytic in the \emph{complex} $g$-plane cut along the negative $g$-axis. There is no rigorous proof of this for general $D$ and represents in $D=1$ a nontrivial result \cite{Martin1969-bk}. 

The CS analysis of the anharmonic oscillator \cite{Collins:1977dw} is expressed in a conventional path integral setting   and has potential implications for higher dimensional  field theory, owing  to the unbounded nature of the potential when $g<0$, a feature common to all $D$ (including $D=0$) \cite{R3b}. For $D=0$ the ``path'' integral defining the integral in the \emph{complex $x$ plane} is given in terms of steepest descent curves with arrows indicating the direction of descent in Fig.\ref{PTsteepestdescent1}. In the model of CS, the  construction of $Z_{0}$ is based on the \cPT -symmetric contours in the \emph{complex $\chi $ plane} in Fig.\ref{PTsteepestdescent2}. These two figures have a striking similarity, but the ``contour" lines in Fig.\ref{PTsteepestdescent2} represent a bundle of functions. It is the existence of a steepest descent ``path'', which goes through both the trivial and nontrivial fixed points (for general $D$), that is the important feature and  leads to the relation 
\be
\label{key}
Z\left( -\lambda \pm i\epsilon; T \right) =Z_{0}\left( -\lambda ;T \right) \pm \frac{1}{2} disc Z\left( -\lambda ;T \right)
\ee
 in the complex $g$-plane, where $\lambda >0$ and $disc$ denotes the imaginary discontinuity across the cut and $Z$ is the partition function in the Hermitian phase. $Z_0$ is the partition function in the \cPT-symmetric phase. As $T \to \infty$, $E\left( g,T \right) \rightarrow E\left( g \right)$ the ground state energy. Hence this approach is useful for understanding ground state properties.
 
 The crux of the method is the ansatz 
 \be
 \label{factorisation}
 \tilde{\phi} \left( \vec x,\tau \right) =\chi \tilde{\phi}_{R} \left( \vec x,\tau \right),
 \ee
where $\tilde{\phi} \left( \vec x,\tau \right)\equiv \lambda^{1/2} \phi \left( \vec x,t \right)$ and $\tilde{\phi}_{R} \left( \vec x,\tau \right)$ is a real field with normalisation $\int dtd{\vec x}\phi_{R}^{2} \left( \vec x,t \right) =1$. By allowing $\chi$ to be a complex variable, CS are able to explore complex deformations of the action as in the $D=0$ case \emph{without} the full apparatus of Lefschetz thimbles. CS considered $D=1$ since this is the relevant case for the quantum mechanical anharmonic oscillator. The complex contours in functional field space are in general very complicated to envisage, but with the CS ansatz the structure is much simplified. In the semiclassical, weak coupling approximation it is only the \emph{immediate neighbourhoods} of the stationary solutions that contribute to the path integral. Since CS can reproduce exactly the classic results in \cite{PhysRevD.7.1620}, this is a sufficient justification for examining its consequences for $D>1$. The relation in \eqref{key} is in terms of quantities which are complicated, particularly for $D>1$, but it represents a \emph{structure} which leads to  the conjectured relation \eqref{ABS}.

 The partition function for $D>1$  is given by 
 \be
 \nonumber
 Z\left( g \right) =\mathcal{N}\int d\left[ \phi \left(\vec x, \tau \right) \right] e^{-S\left[ \phi \right]},
 \ee
 where $\mathcal{N}$ is a normalisation factor,   $\phi \left( \vec x,\tau \right)$ is periodic in $\tau$ with period  $T$, $\vec x$ is the $D-1$ dimensional spatial vector and 
 $$S\left[ \phi \right] =\int_{-T/2}^{T/2} d\tau \  d^{D-1}x\left( \frac{1}{2} \partial^{\mu} \phi \partial_{\mu} \phi +\frac{1}{2} \phi^{2} +\frac{g}{4} \phi^{4} \right).$$

 \noindent The interesting aspect of the approach of \cite{Collins:1977dw}, using (\ref{factorisation}), is that it allows a semiclassical approach similar to the $D=0$ case\cite{R3b,Collins:1977dw}. Hence we shall briefly review the method of steepest descents in the $D=0$ case \cite{R3b}.  
 
 The $D=0$ partition function will be denoted by $z(g)$, to distinguish it from the general $D$ partition function  $ Z\left( g \right)$ . To make a closer connection with field theory, we write the Hermitian case as
\be
z\left( g \right) =\int_{-\infty}^{\infty} d\varphi \  \exp \left[ -\frac{1}{2} \varphi^{2} -\frac{g}{4} \varphi^{4} \right].
\ee
To obtain  $\P$ -symmetric phases, we first write $g=\lambda e^{i\theta}$ (where $\lambda$ is a positive real number)  and  $\tilde{\phi} =\lambda^{1/2} \varphi$; this gives an expression for a contour integral in complex space, which is suitable for a weak coupling semiclassical analysis:
\be
z\left( \lambda e^{i\theta} \right) =\lambda^{-1/2} \int d\tilde{\varphi} \exp \left[ -\frac{1}{\lambda} \tilde{s} \left( \tilde{\varphi} \right) \right]
\ee
where $\tilde{s} \left( \tilde{\varphi} \right) =\frac{1}{2} \tilde{\varphi}^{2} +\frac{1}{4} e^{i\theta}\tilde{\varphi}^{4}$. For convergence of the integral the contour of $\tilde \phi$ has to be rotated by $-\theta/4$ in the complex plane (an example of complexification of the action in field theory). 

For small $\lambda$, the integral can be evaluated by the method of steepest descents \cite{doi:10.1137/1.9780898719260}, where the integral is dominated by contributions at the saddle points $\tilde{\varphi} =0,\  \pm i\sqrt{6} e^{-i\theta /2}$. When $\theta=0$ we have the conventional Hermitian quartic theory and the steepest desecent path goes \emph{only} through the trivial fixed point along the ${\rm Re} \, \phi$  axis (avoiding the fixed points on the imaginary axis).The situation is different for $\theta=\pi$  and the steepest descent path goes through the trivial and   non-trivial fixed points (all on the real axis). There are three such paths, one with \cPT ~symmetry, and two with $\cP$ symmetry \cite{R2}. The \cPT ~phase is constructed with a path ending on hyperbolas in the lower half of the complex $\phi$ plane (see Fig. \ref{PTsteepestdescent1})\cite{R3b}. This contrasts with the $\cP$-symmetric  steepest descent paths, which also go through all the fixed points but pass through opposite quadrants of the complex $\varphi$ plane. The Lefschetz thimbles for $D>1$, are steepest descent paths in complex function space and are very complicated to evaluate. However, in the model of \cite{Collins:1977dw}, fluctuations in the complex degrees of freedom are constrained to be zero dimensional and formally we can extend the analysis to $D>1$. The analogue of the nontrivial saddle points (see Fig. \ref{PTsteepestdescent2}) are the instantons, whose functional form is $D$-dependent (and  are only known numerically for $D>1$ ).

\subsection{ \cb The $D >1$ functional contours \cbl}

We shall follow the methodology used for $D=0$. On writing $g=-\lambda$, we introduce $$\phi \left( \vec x,\tau \right) =\lambda^{-1/2} \tilde{\phi} \left( \vec x,\tau \right).$$ 
In terms of $\tilde{\phi}$
$$S\left[ \phi \right] =\frac{1}{\lambda} \tilde{S} \left[ \tilde{\phi} \right],$$
where 
$$\tilde{S} \left[ \tilde{\phi} \right] =S_{0}\left[ \tilde{\phi} \right]  - 
  S_{I}\left[ \tilde{\phi} \right]= \int_{-T/2}^{T/2} dt\  d^{ D-1}x\left[ \frac{1}{2} \dot{\tilde{\phi}}^{2} +\frac{1}{2} \left( \nabla \tilde{\phi} \right)^{2} +V\left( \tilde{\phi} \right) \right],$$ 
 $V\left( \tilde{\phi} \right) =\frac{1}{2} \tilde{\phi}^{2} -\frac{1}{4} \tilde{\phi}^{4}$,
$$S_{0}\left[ \tilde{\phi} \right] \equiv \int_{-T/2}^{T/2} d\tau \int d^{ D-1}x\left[ \frac{1}{2} \dot{\tilde{\phi}}^{2} +\frac{1}{2}\left( \nabla \tilde{\phi} \right)^{2}+\frac{1}{2} \tilde{\phi}^{2} \right]$$
and 
$$S_{1}\left[ \tilde{\phi} \right] \equiv \int_{-T/2}^{T/2} d\tau \int d^{ D-1}x\;\frac{1}{4} \tilde{\phi}^{4}.$$
Stationary solutions for the action, the analogue of saddle points, are determined by 
\be
\partial^{\mu} \partial_{\mu} \tilde{\phi} =V^{\prime}\left( \tilde{\phi} \right).
\ee
For $D=1$ the solution is $\phi \left( \tau \right) = \sqrt{\frac{2}{3}}\, \sech \,( \tau-\tau_{0})$ and $\tau_0$ is a collective coordinate \footnote{\cb Collective coordinates are parameters that describe the global symmetry preserving motions of these non-perturbative field configurations and need to be integrated over to give the total contribution of the instantons.\cbl}, but for $D>1$ can be determined numerically or obtained approximately using matched asymptotics. For all $D$ there is the trivial fixed point $\tilde{\phi}=0$ and a continuum (labelled by collective coordinates) of nontrival saddle points. 

In the functional integral there are integration contours in the complex $\chi$ plane as well as functional integrals over real fields. For $\theta \neq 0$, $\tilde{\phi}_R$ remains real and ${\rm arg} \chi =-\theta /4$, just as in the $D=0$ case. For $\theta =\pi$, the $\chi$ contour is a straight line through the origin in the upper right quadrant of the complex $\chi$ plane at an angle $\pi/4$. It is shown in \cite{Collins:1977dw} that the location of the saddle points in the $\chi$ plane  are at 
\begin{align}
\label{CSsaddlepoint}
 \chi& =0 \quad {\rm and} \nonumber \\  
 \chi_{R}\left[ \tilde{\phi}_{r} \right] &=\left( \frac{1}{2} \frac{S_{0}\left[ \tilde{\phi}_{r} \right]}{S_{I}\left[ \tilde{\phi}_{r} \right]} \right)^{1/2},
\end{align}

\noindent for any $\tilde{\phi}_{r}$. $\chi_{R}\left[ \tilde{\phi}_{r} \right]$  is
 a homogeneous function of degree $-1$. Independent of ${\tilde{\phi}}_{r}$ the leading behaviour 
is controlled by the trivial fixed point. Hence the leading behaviour of \emph{both} 
$Z\left[ \lambda e^{i\pi} \right]$ and $Z\left[ \lambda e^{-i\pi} \right]$ and their sum are controlled by the trivial fixed point. 
Hence ${Z_{0}}(-\lambda)$ is given  by weak coupling perturbation theory and is real. By analytic continuation in the $\chi$ plane, from the contour ${Z_{0}}(-\lambda)$ can be identified with the \cPT ~symmetric contribution, which is real. The imaginary parts of 
$Z \left[ \lambda e^{ i\pi} \right]$ are related to the nontrivial fixed points, which give rise to nonanayltic instanton contributions. These arguments are possible because of the ansatz \eqref{factorisation}. A fixed point represents a continuum of stationary functions (solutions of the equations of motion) \cite{Collins:1977dw}. The $\chi$ part of the contour follows those in the $D=0$ case, but the lines emanating from fixed points represent real functions, which are complexified through $\chi$. Consequently, within this framework, we can derive
\be
\label{ABS}
Z^{PT}\left( \lambda \right) =\frac{1}{2} \left( Z_{Herm}(-\lambda +i0+ \right) +Z_{Herm}\left( -\lambda -i0+ \right) ,
\ee
a version of the Ai-Bender-Sarkar conjecture. However this framework relies on the ansatz \eqref{factorisation}, which simplifies the much more complicated features in a Lefschetz thimble discussion (that remains  a difficult unsolved problem).

%%%%%%%%%%%%%%%%%%%%%%%%%%%%%%
\cbl

\section{Conclusions} 
The connection between physical quantities for scalar theories \cb with quartic interaction \cbl in their  parity-symmetric   and  \cPT-symmetric phases involves a continuation in the coupling which is  non-analytic. For single component scalar theories with a quartic interaction in $D=1$, the relation conjectured in \cite{R3b} holds for ground and excited states in \emph{weak} coupling, according to our numerical calculations, but not for strong coupling.   For $D=0$ the conjecture is in terms of partition functions and holds for many cases, but  for multi-component fields \cb the conjecture does not hold in general. \cb Furthermore, for $D=1$ the $O(N)$ symmetric theories do not satisfy the conjecture. For the $D>1$ theory of the single-component scalar, in an \emph{approximate} treatment of Lefschetz thimbles, the conjecture for partition functions is satisfied. This is only a plausibility argument  based on the ansatz of CS \cite{Collins:1977dw}.
Whether relations of this type hold for quantities such as beta functions and partition functions in a more complete treatment of Lefschetz thimbles (for $D>1$) is currently not known \cite{Lawrence:2023woz}. \cbl ~It is possible that renormalisation group analysis, and the role of Landau poles \cite{Romatschke:2022llf,Mavromatos:2024ozk}, may lead to connections between Hermitian and \cPT-symmetric theories.  Such an analysis is used recently in an attempt to understand the role of dark energy at large length scales, based on Chern-Simons gravity \cite{Mavromatos:2024ozk} modified by \cPT~ symmetry.

\vskip 2cm

\section*{Acknowledgments}
We would like to thank Wen-Yuan Ai, Carl Bender, Daniele Dorigoni and Syo Kamata for discussions on the phase structure of the quantum anharmonic oscillator and transseries.  L.C. is supported by the King’s-China Scholarship Council. The work of S.S. is supported  by  the EPSRC grant EP/V002821/1. 

\begin{appendix}
\section{Numerical method of obtaining energy eigenvalues of a Schr\"odinger equation along the \cPT contour}
\label{RungeKutta}

This appendix outlines a modified Runge-Kutta method \cite{Suli2003-hd} for 
obtaining the energy spectrum of a Schr\"odinger equation with the potential of the form
\begin{equation}
    V (x) = \frac{1}{2} m^2 x^2 + \frac{1}{4} g x^2 (ix)^\delta.
\end{equation}
The method is based on a procedure for the continuation of eigenvalue problems \cite{Bender1993-hk}. It is used in \cite{R1}, where details of the numerical method are not given. We are primarily interested in $\delta \to 2$.
Since the method is not commonly used outside the study of  \cPT ~symmetry, in order to make this work more  accessible, we  give some details of a modified shooting method  \footnote{We implement numerical methods  using Python 3.}. It is necessary to continue $x$ into the complex plane in this method.

\subsection{Stokes wedges in the complex plane and shooting methods}
\label{Stokes}

We normally solve a Schr\"odinger equation in the domain $-\infty <x<\infty$ and boundary condition for the wavefunction $|\psi \left( x \right) |\rightarrow 0$ as $|x|\rightarrow \infty$. The possible need to deform $x$ into the complex plane for $\delta \ne 0$ can be seen from  WKB analysis (in the geometric optics approximation) \cite{R15} and the condition for the wavefunction to be normalisable. From WKB analysis $\psi \left( x \right) \sim \exp \left[ \pm \int_{}^{x} \sqrt{Q\left( s \right)} ds \right]$ where $Q\left( s \right) =E-V\left( s \right)$. The integration path for $x$ in the integral is along a contour in the complex plane
$$s\left( \tau \right) =r\left( \tau \right) \exp \left( i\theta \left( \tau \right) \right),$$ where $\tau$ parametrises the path  and $r\left( \tau \right) \to \infty$ as $\tau \to \infty$. For large $\left| x \right|$  
$$\psi \left( x \right) \sim \exp \left[ \pm \frac{2}{4+\delta} i^{\delta /2}x^{2+\delta /2} \right].$$
We deduce that $\psi \left( x \right) \to 0$ sufficiently fast as $r \to \infty$ in a Stokes sector (wedge) $\theta_{1} <\theta <\theta_{2}$, which depends on $\delta$. The centre angle of the sector  is given by $\theta =\left( \theta_{1} +\theta_{2} \right) /2$. For the decaying solution there is a left sector and a right sector with centre angles $\theta_{L}$ and $\theta_{R}$: $\theta_{L} =-\pi +\frac{\delta}{\delta +4} \  \frac{\pi}{2}$  and $\theta_{R} =-\frac{\delta}{\delta +4} \  \frac{\pi}{2}$. The opening angle of a wedge is $\frac{2\pi}{\delta +4}$.

\begin{figure}[htbp]
    \centering
\begin{tikzpicture}
    \draw[gray, thick] (-5,0) -- (5,0) node[anchor=north]{$\text{Re}x$} ;
\draw[gray, thick] (0,-3) -- (0,1) node[anchor=west]{$\text{Im}x$};
\draw[black,  dashed] (-4,-2) -- (0,0) node[anchor=north east]{ $\theta_{L}\quad \quad$       };
\draw[black, dashed](4,-2)   -- (0,0) node[anchor=north west]{$\quad \quad\theta_{R}$};
\end{tikzpicture}    
\caption{Centres of the Stokes wedges with  angles of $\theta_{L}$ and $\theta_{R}$.}
\label{Stokes coutour}
\end{figure}
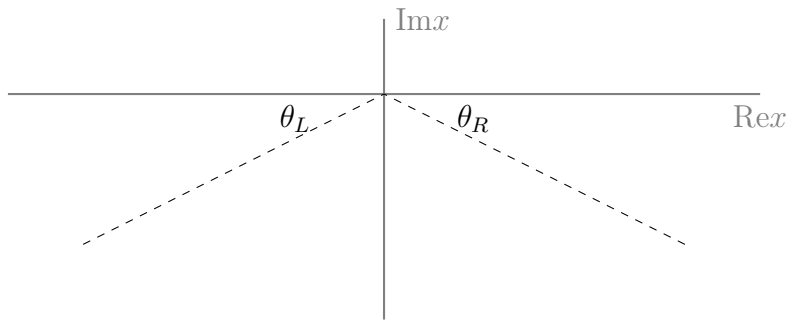

\noindent For systems which are $\cP$ symmetric, parity even and odd solutions are selected using $\psi^{\prime} \left( 0 \right) =0$ and $\psi\left( 0 \right) =0$ respectively. For the \cPT-symmetric case, these conditions could be replaced by $\psi_{L} \left( 0 \right) =\psi_{R} \left( 0 \right)$. Because of the linearity of the Schr\"odinger equation, linear superpositions of solutions remain solutions. Maintaining this property, when shooting from the left and right sectors, is not guaranteed. This issue is addressed \cite{R1} through an altered condition at $x=0$
\be
\label{shooting condition}
\frac{\psi_{L}^{\prime} \left( x \right)}{\psi_{L} \left( x \right)} -
\frac{\psi_{R}^{\prime} \left( x \right)}{\psi_{R} \left( x \right)} =0.
\ee
The standard shooting method is now modified by taking initial guesses in the left sector and right sector and integrating towards $x=0$. A Runge-Kutta method is viable because the parametrisation of the path guarantees that we are dealing with a one variable problem.

%%%%%%%%%%%%%%%%%%%%%%%%%%%%%%
\vskip .5cm
\subsection{Algorithm of the modified Runge-Kutta method}
Our Python program needs  the input arguments:

\begin{enumerate}
    \item The value of $m$ and $g$: these are the coupling parameters for the $x^2$ term and the $x^4$ term respectively.
    \item The value of $\delta$: this is the power index in $x^2 (ix)^\delta$. We take $\delta = 2$. As we have noted $\delta$  plays an important role in determining the \cPT ~contour. 
    \item Two estimated values of theeigenvalue, namely $E_1$ and $E_2$.  This program does not produce the entire energy spectrum at once, instead, one first need to guess some value for the eigenvalue and run the program. After some iterations, $E_1$ and $E_2$ will gradually converge to the closest eigenvalue.
    \item The specification of the range of integration, $[-r_0,r_0]$. In order to calculate the eigenvalue, one needs to numerically obtain the functions. This is done by integrating from both sides of the contour towards $r=0$. The greater the chosen value of the initial position $r_0$, the longer the calculation will take. 
\end{enumerate}

We deform the integration contour from the real axis into the complex plane. For $\delta=2$, the centre angles of the Stokes wedges are  $\theta_{L} =-\pi +\frac{\pi}{6}$  and $\theta_{R} =- \frac{\pi}{6}$. We write $x = r e^{i\theta}$  to go from the real axis to the complex plane. For constant $\theta$, $\frac{\mathrm{d}}{\mathrm{~d} x}= e^{-i\theta}\frac{\mathrm{d}}{\mathrm{~d} r}$. Then our second order differential equation
\begin{equation}
    \frac{\mathrm{d}^2}{\mathrm{~d} x^2} \psi (x) = \left(\frac{1}{2} m^2 x^2 + \frac{1}{4} g x^2 (ix)^\delta -E \right)\psi (x)
\end{equation}
becomes
\begin{equation}
    \frac{\mathrm{d}^2}{\mathrm{~d} r^2} \psi (r) = e^{i2\theta} \left(\frac{1}{2} m^2 r^2 e^{i2\theta}+ \frac{1}{4} g r^2 e^{i2\theta} (ir e^{i\theta})^\delta -E \right)\psi (r).
\end{equation}
We need to integrate numerically the function $\psi(r)$  from both sides to the origin. We set the initial value of $\psi$ and $\psi'$ at both ends of the range of integration to be $1$, since these are free parameters whose choice does not affect the final result. \footnote{The solutions of a second order differential equation would naturally have two degrees of freedom, so the initial conditions would not affect the calculations for eigenvalues as long as these initial conditions are kept the same on both sides.} For our differential equation, We take all the initial conditions to perform numerical integration from both sides of the origin. 

Next we  illustrate algorithm of the Runge-Kutta method as applied to our problem. 
Because the Runge-Kutta method is designed to solve a set of coupled first order differential equation,  we write our second order differential equation as two first order differential equations, denoted by $F$ and $G$: 
\begin{equation}
    \begin{aligned}
      F\equiv  \frac{\mathrm{d}}{\mathrm{~d} r} \psi (r) &=   a(r), \\
       G\equiv \frac{\mathrm{d}}{\mathrm{~d} r} a(r) &=  e^{i2\theta} \left(\frac{1}{2} m^2 r^2 e^{i2\theta} + \frac{1}{4} g r^2 e^{i2\theta} (ir e^{i\theta})^\delta -E \right)\psi (r).
    \end{aligned}
\end{equation}
In our case, the functions $F$ and $G$ require the following 5 arguments: $r,\theta,\psi,\psi',E$. We see that $F$ only depends on the first derivative of $\psi$, while $G$ depends on all other arguments through the  second order differential equation that we wish to solve. 

For some small step size $h$ (not to be confused with $h$ in \eqref{muller}, the value of $\psi$ and $\psi'$ at $r+h$ can be approximated by
\begin{equation}
    \begin{aligned}
        \psi(r+h) &= \psi(r) + \frac{1}{6}(k_1+ 2k_2 + 2k_3 + k_4), \\
         \psi'(r+h) &= \psi'(r) + \frac{1}{6}(l_1+ 2l_2 + 2l_3 + l_4),
    \end{aligned}
\end{equation}
where $k_i$ and $l_i$ are given by
\begin{equation}
    \begin{aligned}
         k_1 &= h \times F(r, \theta, \psi, \psi', E), \\
        l_1 &= h \times G(r, \theta, \psi, \psi', E),\\
        k_2 &= h \times F(r + \frac{1}{2}h , \theta, \psi +\frac{1}{2} k_1 , \psi' + \frac{1}{2}l_1 , E),\\
        l_2 &= h \times G(r + \frac{1}{2}h, \theta, \psi +\frac{1}{2} k_1 , \psi' + \frac{1}{2}l_1 , E),\\
        k_3 &= h \times F(r + \frac{1}{2}h , \theta, \psi + \frac{1}{2}k_2 , \psi' +\frac{1}{2} l_2 , E),\\
        l_3 &= h \times G(r + \frac{1}{2}h , \theta, \psi +\frac{1}{2} k_2 , \psi' +\frac{1}{2} l_2 , E),\\
        k_4 &= h \times F(r + h, \theta, \psi + k_3, \psi' + l_3, E),\\
        l_4 &= h \times G(r + h, \theta, \psi + k_3, \psi' + l_3, E).\\
    \end{aligned}
\end{equation}
($h$  can be altered to control the precision of the estimation.) The smaller the $h$ one chooses, the more steps one needs to take to reach the origin from  a point $x$. The result can be  more accurate, but is also more time consuming. Note that because we are integrating inwards, $h$ is a negative number in our case\footnote{
A possible step size would be $h= -0.0005$ for the purpose of obtaining data for this paper}. Thus the total number of steps to go from a point labelled by $r$ to the origin would be $\frac{r}{\left| h \right|}$. 

We need to do this process from both sides of the origin; since we  deform our contour into the complex plane, the only difference between the left and right side of the complex $x$-plane,  is the choice of  angle $\theta$, $\theta_L$ for  left  and $\theta_R$ for the right side. So we obtain two values for $\psi$, namely $\psi_L$ and $\psi_R$, and two values for $\psi'$, namely $\psi'_L$ and $\psi'_R$. Since the differential equation is invariant under Parity, we  normally expect either the wave function or the first derivative of the wave function to vanish at the origin. But, because we  deform our contour, we  require
\begin{equation}
\label{shooting condition 2}
    e^{-i\theta_L}\frac{\psi_{L}^{\prime} \left( r \right)}{\psi_{L} \left( r \right)} -e^{-i\theta_R}\frac{\psi_{R}^{\prime} \left( r \right)}{\psi_{R} \left( r \right)} = 0.
\end{equation}
This condition comes from equation \eqref{shooting condition} with a change of variable since $\psi_{L}^{\prime} \left( r \right) = \frac{\mathrm{d}}{\mathrm{~d} r} \psi (r) = \frac{\mathrm{d}x}{\mathrm{~d} r}\frac{\mathrm{d}}{\mathrm{~d} x} \psi (x) =   e^{i\theta_L} \frac{\mathrm{d}}{\mathrm{~d} x} \psi (x) $ and similarly for $\psi_{R}^{\prime} $.

Ideally, the shooting condition \eqref{shooting condition 2} should be automatically satisfied if the value of $E$ is an eigenvalue of the differential equation. However, since the eigenvalue is the quantity we wish to calculate, the chance of any of our guessed values $E_1$ and $E_2$ being the eigenvalue is  small. Hence in practice, we  end up with
 \begin{equation}
   A(E) =  e^{-i\theta_L}\frac{\psi_{L}^{\prime} \left( r \right)}{\psi_{L} \left( r \right)} -e^{-i\theta_R}\frac{\psi_{R}^{\prime} \left( r \right)}{\psi_{R} \left( r \right)} \neq 0
\end{equation}
We note that the smaller $A(E)$ is, the closer this value of $E$ is to the actual eigenvalue. Hence we  compare the value of $A(E_1) $ and $A(E_2)$ to see which one is smaller, to determine whether  $E_1$ or $E_2$ is closer to the eigenvalue. \cbl We  replace the one that is further from the actual energy by a better guess \cbl
\begin{equation}
    E_3 = \frac{E_1 A(E_2)-E_2 A(E_1)}{A(E_2)-A(E_1)}.
\end{equation}
This process  minimises the value of $A$ and leads  to a better approximation of the energy. One can perform this process as many times as required to obtain a sequence of values which converge to the actual eigenvalue (up to a certain precision).

\section{\cb Steepest descent paths and stationary points \cbl}
\label{Steepestdescent}

\begin{figure}[htbp]
  \centering
\begin{tikzpicture}[scale=1] 
    \draw[->,black,  line width=0.5mm] (-7,0) -- (7,0) node[anchor=north]{\large $\text{Re}x$} ;
\draw[->,black,  line width=0.5mm] (0,-7) -- (0,7) node[anchor=west]{\large $\text{Im}x$};
\draw[black ,  dashed] (-6.5,6) ..controls (-3.5,3.5) and (-2.7,2).. (-2.5,0) ;
\draw[black ,  dashed] (6.5,6) ..controls (3.5,3.5)and (2.7,2).. (2.5,0) ;
\draw[<-,red, line width=0.7mm] (-6.5,-6) ..controls (-3.5,-3.5) and (-2.7,-2).. (-2.5,0) ;
\draw[<-,red, line width=0.7mm] (6.5,-6) ..controls (3.5,-3.5)and (2.7,-2).. (2.5,0) ;
\node at (-3,-0.3) {\large $x_L$};
\node at (3,-0.3) {\large $x_R$};
\node at (0.5,0.3) {\large $x_0$};
\draw[->,red, line width=0.7mm](0,0) -- (-2.5,0);
\draw[->,red, line width=0.7mm](0,0) -- (2.5,0);

\end{tikzpicture}    
\caption{$D =0 $ case (similar to Figure 2 in Ref.\cite{R3b}) where $x_L,x_R$ are the nontrivial saddle points of $V_2$ }
\label{PTsteepestdescent1}
\end{figure}
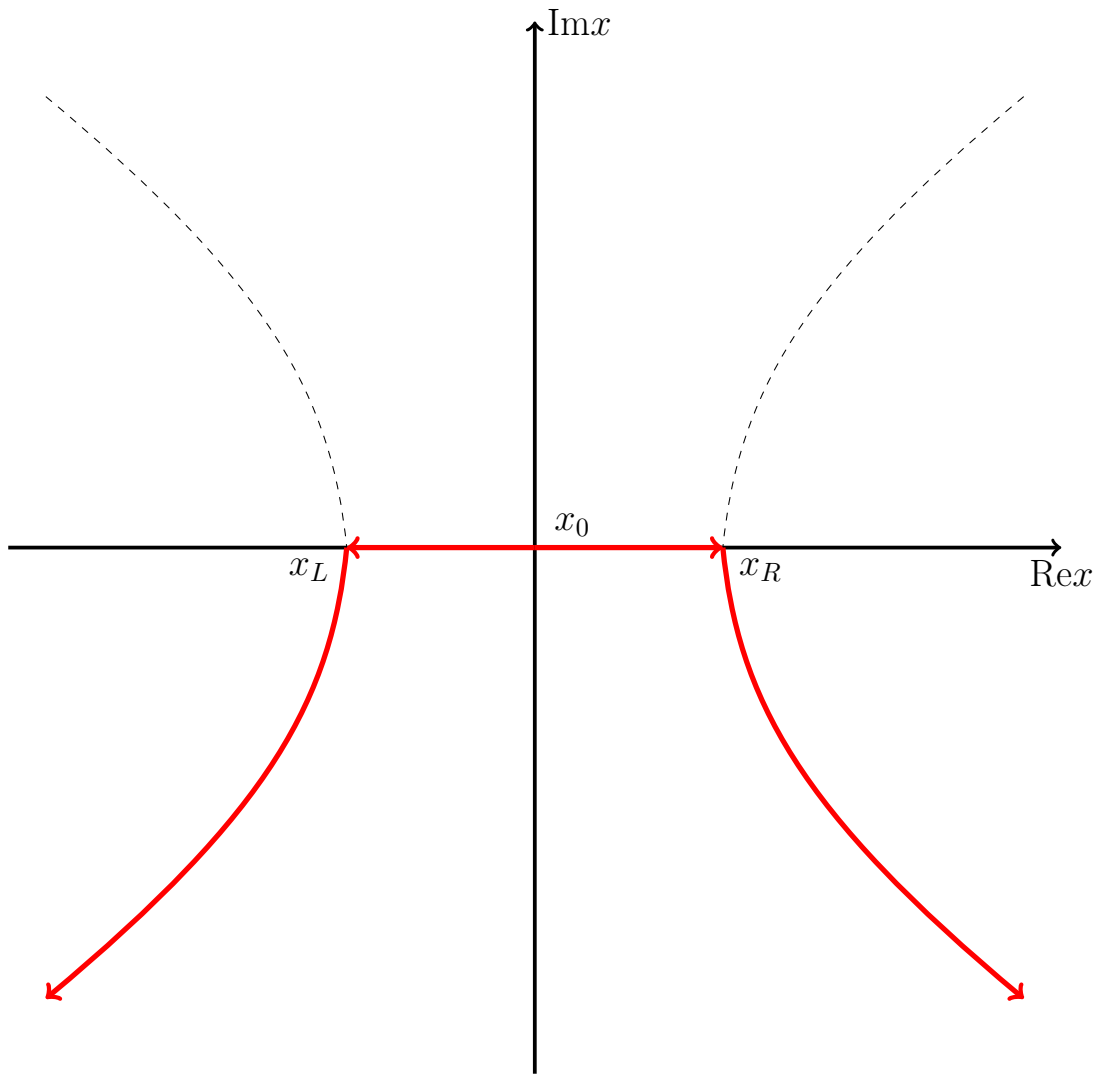
\begin{figure}[htbp]
    \centering
\begin{tikzpicture}[scale=1] 
    \draw[->,black,  line width=0.5mm] (-7,0) -- (7,0) node[anchor=north]{\large $\text{Re}\chi$} ;
\draw[->,black,  line width=0.5mm] (0,-7) -- (0,7) node[anchor=west]{\large $\text{Im}\chi$};
\draw[black ,  dashed] (7,7) -- (4,4) ;
\draw[black ,  dashed]  (4,4) -- (4,0) ;
\draw[->, line width=0.7mm, red]  (0,0) -- (4,0) ;
\draw[->, line width=0.7mm, red]  (4,0) -- (4,-4) ;
\draw[->, line width=0.7mm, red](4,-4) -- (7,-7);
\draw[black ,  dashed] (-7,7) -- (-4,4) ;
\draw[black ,  dashed]  (-4,4) -- (-4,-0) ;
\draw[->, line width=0.7mm, red]  (0,0) -- (-4,0) ;
\draw[->, line width=0.7mm, red]  (-4,0) -- (-4,-4) ;
\draw[->, line width=0.7mm, red](-4,-4) -- (-7,-7);
\node at (3.5,0.5) {\large $\chi_R$};
\node at (-3.5,0.5) {\large $\chi_L$};
\end{tikzpicture}    
\caption{Corresponding $D\geq1$ case, adapted from Ref. \cite{Collins:1977dw}, and $\chi_R$ ($=-\chi_L$)  defined in \eqref{CSsaddlepoint}}
\label{PTsteepestdescent2}
\end{figure}
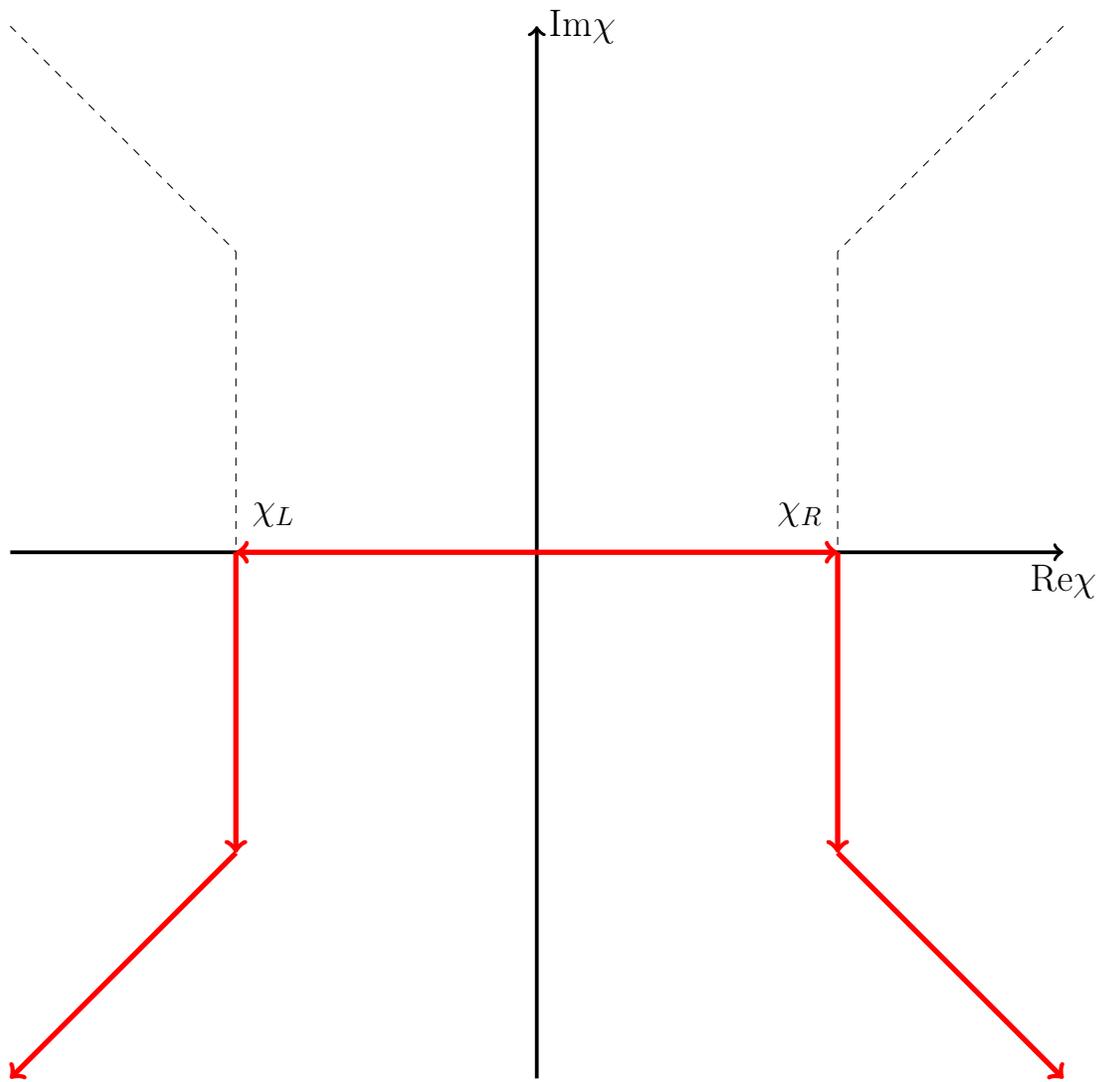

\end{appendix}
\vskip 15cm
%%%%%%%%%%%%%%%%%%%%%%%%%%%%%%%%%%%%%%%%%%%%%%%%%%%%%
\bibliographystyle{utphys}
\bibliography{source.bib}

\end{document}